\documentclass[twocolumn]{aastex63}
\usepackage[figuresright]{rotating}
\usepackage{subfigure} 
\DeclareGraphicsExtensions{.pdf,.jpg,.png}
\usepackage{float}

\usepackage{threeparttable}

\begin{document}


\title{The Faraday rotation measure of the M87 jet at 3.5mm with the Atacama Large Millimeter/submillimeter Array}
\correspondingauthor{Sijia Peng, Ru-Sen Lu}
\email{sjpeng@shao.ac.cn, rslu@shao.ac.cn}


\author[0000-0001-8492-892X]{Sijia Peng}
\affiliation{Shanghai Astronomical Observatory, Chinese Academy of Sciences, Shanghai 200030, P. R. China}
\affiliation{School of Astronomy and Space Science, Nanjing University, Nanjing 210023, P. R. China}
\affiliation{Key Laboratory of Modern Astronomy and Astrophysics, Nanjing University, Nanjing 210023,  P. R. China}

\author[0000-0002-7692-7967]{Ru-Sen Lu}
\affiliation{Shanghai Astronomical Observatory, Chinese Academy of Sciences, Shanghai 200030,  P. R. China}
\affiliation{Key Laboratory of Radio Astronomy and Technology, Chinese Academy of Sciences, A20 Datun Road, Chaoyang District, Beijing, 100101, P. R. China}
\affiliation{Max-Planck-Institut für Radioastronomie, Auf dem Hügel 69, D-53121 Bonn, Germany}

\author[0000-0002-2542-7743]{Ciriaco Goddi} 
\affil{
Universidade de São Paulo, Instituto de Astronomia, Geofísica e Ciências Atmosféricas, Departamento de Astronomia, São Paulo, SP 05508-090, Brazil}
\affiliation{Dipartimento di Fisica, Universit\'a degli Studi di Cagliari, SP Monserrato-Sestu km 0.7, I-09042 Monserrato,  Italy}
\affiliation{INAF - Osservatorio Astronomico di Cagliari, via della Scienza 5, I-09047 Selargius (CA), Italy}
\affil{
INFN, Sezione di Cagliari, Cittadella Univ., I-09042 Monserrato (CA), Italy}

\author[0000-0002-4892-9586]{Thomas P. Krichbaum}
\affiliation{Max-Planck-Institut für Radioastronomie, Auf dem Hügel 69, D-53121 Bonn, Germany}

\author[0000-0003-0355-6437]{Zhiyuan Li}
\affiliation{School of Astronomy and Space Science, Nanjing University, Nanjing 210023, P. R. China}
\affiliation{Key Laboratory of Modern Astronomy and Astrophysics, Nanjing University, Nanjing 210023,  P. R. China}

\author[0000-0003-1576-0961]{Ruo-Yu Liu}
\affiliation{School of Astronomy and Space Science, Nanjing University, Nanjing 210023,  P. R. China}
\affiliation{Key Laboratory of Modern Astronomy and Astrophysics, Nanjing University, Nanjing 210023,  P. R. China}

\author[0000-0001-8229-7183]{Jae-Young Kim}
\affiliation{Department of Astronomy and Atmospheric Sciences, Kyungpook National University, Daegu 702-701, Republic of Korea}
\affiliation{Max-Planck-Institut für Radioastronomie, Auf dem Hügel 69, D-53121 Bonn, Germany}

\author[0000-0001-6081-2420]{Masanori Nakamura}
\affiliation{National Institute of Technology, Hachinohe College, 16-1 Uwanotai, Tamonoki, Hachinohe City, Aomori 039-1192, Japan}
\affiliation{Institute of Astronomy and Astrophysics, Academia Sinica, 11F of Astronomy-Mathematics Building, AS/NTU No. 1, Sec. 4, Roosevelt Road, Taipei 10617, Taiwan, R.O.C.}

\author[0000-0003-3564-6437]{Feng Yuan}
\affiliation{Center for Astronomy and Astrophysics and Department of Physics, Fudan University, Shanghai 200438, P. R. China}

\author[0000-0002-1908-0536]{Liang Chen}
\affiliation{Shanghai Astronomical Observatory, Chinese Academy of Sciences, Shanghai 200030,  P. R. China}
\affiliation{Key Laboratory for Research in Galaxies and Cosmology, Chinese Academy of Sciences, P. R. China}

\author[0000-0003-3708-9611]{Iván Martí-Vidal}
\affiliation{Departament d’Astronomia i Astrofísica, Universitat de València, C. Dr. Moliner 50, 46100 Burjassot, València, Spain}
\affiliation{Observatori Astronòmic, Universitat de València, C. Catedrático
José Beltrán 2, 46980 Paterna, València, Spain}

\author[0000-0003-3540-8746]{Zhiqiang Shen}
\affiliation{Shanghai Astronomical Observatory, Chinese Academy of Sciences, Shanghai 200030,  P. R. China}
\affiliation{Key Laboratory of Radio Astronomy and Technology, Chinese Academy of Sciences, A20 Datun Road, Chaoyang District, Beijing, 100101, P. R. China}




\begin{abstract}
Faraday rotation is an important probe of the magnetic fields and magnetized plasma around active galactic nuclei (AGN) jets. We present a Faraday rotation measure image of the M87 jet between 85.2\,GHz and 101.3\,GHz with a resolution of $\sim 2\arcsec$ with the Atacama Large Millimeter/submillimeter Array (ALMA). We found that the rotation measure (RM) of the M87 core is $\rm (4.5\pm 0.4)\times10^{4}\ rad\ m^{-2}$ with a low linear polarization fraction of $\rm (0.88\pm 0.08)\%$. 
The spatial RM gradient in the M87 jet spans a wide range from $\sim -2\times10^4\rm~rad\ m^{-2}$ to $\sim 3\times10^4\rm~rad\ m^{-2}$ with a typical uncertainty of $0.3\times10^4\rm~rad\ m^{-2}$. A comparison with previous RM measurements of the core suggests that the Faraday rotation of the core may originate very close to the super massive black hole (SMBH). Both an internal origin and an external screen with a rapidly varying emitting source could be possible.
As for the jet, the RM gradient indicates a helical configuration of the magnetic field that persists up to kpc scale. Combined with the kpc-scale RM measurements at lower frequencies, we found that RM is frequency-dependent in the jet. One possible scenario to explain this dependence is that the kpc-scale jet has a trumpet-like shape and the jet coil unwinds near its end.
\end{abstract}




\section{Introduction}

Magnetic fields are believed to play a vital role in the formation of relativistic jets \citep{2019ARA&A..57..467B}, either by extracting energy from a spinning supermassive black hole (SMBH) via the Blandford-Znajek mechanism \citep{1977MNRAS.179..433B}, or by tapping the rotational energy of a magnetized accretion flow, known as the Blandford-Payne mechanism \citep{1982MNRAS.199..883B}. 
The detailed information of magnetic field strength and topology can be obtained by polarization observations \citep[e.g.,][]{2011MNRAS.415.2081C}.

Faraday rotation of the polarization position angle occurs when the linearly polarized radiation travels through a magnetized plasma   \citep{1966MNRAS.133...67B}. The amount of Faraday rotation is proportional to free electron density $n_{e}$ (in $\rm cm^{-3}$) along the propagation path and the magnetic field component ($\rm B_{\parallel}$, unit in $\mu G$) along the line of sight (with the path length $l$ in $pc$). The Faraday depth ($\rm \phi(l)$) is defined as
\begin{equation}
\phi(l)= 0.81 \int^{d}_{0} n_{e}\ B_{\parallel} \cdot dl\ \mathrm{rad\ m^{-2}}. 
\end{equation}
Conventionally, a positive Faraday depth means that the magnetic field is pointing toward the observer. If there is only one source along the line of sight and it has no internal Faraday rotation, the observed Faraday depth would be equal to its Faraday rotation measure (RM).
In this simplest case, the observed electron vector position angle (EVPA, $\chi_{\mathrm{obs}}$) of the target has a linear dependence on the square of the wavelength ($\rm \lambda^{2}$), 
\begin{equation}
\chi_{\mathrm{obs}}= \chi_{0}+RM \lambda^{2} \label{eqn:RM}.
\end{equation}
 where $\chi_{0}$ is the intrinsic EVPA of the emission region. We can derive the RM by measuring the observed EVPA of the target at multiple wavelengths assuming that the opacity is the same at these wavelengths.

The Faraday RM sign reversal spatially occurring across a jet has been found in a number of studies that suggest the existence of a helical magnetic field structure in a relativistic jet, such as the case of 3C273 \citep[e.g.,][]{2002PASJ...54L..39A}. As a well-known target for studying SMBH physics, M87 \citep[$D \rm \approx  16.7\pm 0.9\ Mpc$, ][]{2009ApJ...694..556B,2010A&A...524A..71B} launches a kpc-scale FR-I type relativistic jet \citep{1980ApJ...239L..11O}. 
Radio observations over the past two decades have intensively studied its jet from the arcsecond \citep[e.g.,][]{1990ApJ...362..449O,2002ApJ...564..683M,2013ApJ...774L..21M,2014ApJ...783L..33K,2016ApJ...832....3A,2021ApJ...910L..14G} to milli-arcsecond (mas) scales \citep[e.g.,][]{1999Natur.401..891J,2007ApJ...668L..27K,2012ApJ...745L..28A,2016ApJ...817..131H,2018A&A...616A.188K,2018ApJ...855..128W,2019ApJ...871..257P,2023Natur.616..686L,2023Natur.621..711C}.  
However, robust evidence for the existence of an ordered large-scale helical magnetic field structure in its jet is still lacking. 
Using $Hubble\ Space\ Telescope $ (HST) polarimetric imaging and Very Large Array (VLA) observations at $8-43$ GHz, \citet{2016ApJ...832....3A} found that an apparent helical structure in the EVPA exists in several jet knots on kpc scale. However, \citet{2016ApJ...823...86A} pointed out that no significant RM gradient was found in the kpc-scale jet based on their VLA observations, especially for a sign reversal of the transverse RM.
Recently, \citet{2021ApJ...923L...5P} reported a reversal of the RM sign across the M87 jet in konts E and F across the M87 jet from VLA observations at $4-18$\,GHz.

Polarization observations at millimeter wavelengths can also detect possible RM gradients. Due to less (differential) Faraday and opacity effects, a radio jet is usually more polarized at millimeter wavelengths \citep{2010A&A...515A..40T,2014JKAS...47...15T}. Here, we present the first Faraday RM image of the M87 jet from 3.5\,mm (Band 3) Atacama Large Millimeter/submillimeter Array (ALMA) observations.
In Section 2, we describe our ALMA observations and data reduction. In Section 3, we show the polarization results and the RM variations along the jet. The analysis of the RM of the unresolved core and the kpc-scale jet is discussed in Section 4. We summarize our results in Section 5.

\section{Observation and data reduction}
{\label{sec:data}}
Our Cycle 5 ALMA observations of M87 (Project Code: 2017.1.00842.V) were performed on April 14–15, 2018, with 32 antennas of 12 meter diameter at Band 3 using the C43-3 configuration. The maximum baseline length is 500 m and the minimum length is 15 m.
These observations cover 85.3 GHz to 101.3 GHz with four full-polarization subbands (spectral windows, SPWs) centered at 86.3, 88.3, 98.3, and 100.3\,GHz with a bandwidth of 1.875\,GHz each (corresponding to a maximum detectable RM of $\rm 1.73 \times 10^6\rm~rad\ m^{-2}$ and a minimum of $(4.8-19) \times 10^3\rm~rad\ m^{-2}$). The ALMA observations were performed in phased-array mode \citep{2018PASP..130a5002M,2023PASP..135b5002C}
as part of a Global Millimeter VLBI Array (GMVA) imaging experiment \citep{2023Natur.616..686L}. 3C\,273 and 3C\,279 were observed in interleaved scans as calibrators. The observation lasted about 8.0 hours and the total on-source time for M87 was about 1.5 hours.

\begin{deluxetable*}{ccccccc}
\tablecaption{Polarization properties of M87 core and the selected jet knots at 86.3 GHz (SPW0) in April 2018 \label{tab:knot}\\ 
}
\tablecolumns{6}
\tablenum{1}
\tablewidth{0pt}
\tablehead{
\colhead{Position}  &\multicolumn{2}{c}{Coordinate }& \colhead{$I_{\rm Stokes I,peak}$ } & \colhead{$I_{\rm pol,peak}$} &\colhead{$I_{\rm pol,tot}$} & \colhead{EVPA} \\
\colhead{} & \colhead{R.A.} & \colhead{Decl.} & \colhead{($\rm mJy\ beam^{-1}$)} & \colhead{($\rm mJy\ beam^{-1}$)} & \colhead{($\rm mJy$)} &\colhead{($\rm \arcdeg) $}
}
\colnumbers  
\startdata
Core & 12h30m49.423s &  +12d23m28.05s & $\rm 1422\pm10$ & $\rm 8.3\pm 0.7$  & $\rm 11.3\pm 1.3$  & $\rm 11.7\pm3.1$ \\   
Knot A & 12h30m48.630s & +12d23m32.50s & $\rm 233\pm 18$ &$\rm 18.4\pm1.2 $  & $\rm 29.4\pm2.9 $  & $\rm -27.8\pm4.8$ \\ 
Knot B & 12h30m48.474s & +12d23m33.04s & $\rm 209\pm 19$ &$\rm 40.6\pm2.0$  & $\rm 55.1\pm3.4$ & $\rm 33.0\pm3.2 $ \\ 
Knot C & 12h30m48.146s & +12d23m35.84s & $\rm 87\pm 8$ &$\rm 17.7\pm1.3$  & $\rm 39.4\pm3.9$ & $\rm -34.0\pm3.7 $ \\
Knot D & 12h30m49.173s & +12d23m29.30s & $\rm 33\pm 3$ &$\rm 9.6\pm0.4$  & $\rm 17.4\pm0.8$ & $\rm 16.8\pm2.7 $ \\ 
Knot F &  12h30m48.837s & +12d23m31.34s & $\rm 53\pm5$ & $\rm 11.1\pm0.4 $ & $\rm 24.3\pm1.1$ & $\rm 17.6\pm4.1 $ 
\enddata
\tablecomments{(1) Selected positions measured by IMFIT. (2) - (3) J2000 coordinates of the fitted knot centroid. (4) The Stokes I peak flux density. (5) The peak polarized flux density. (6) The integrated polarized flux density. (7) The average electron vector position angle (EVPA) in four SPWs at the centroid. The EVPA errors includes a 5\% systematic error in absolute flux-scale calibration and an additional 0.03\% systematic error due to the leakage from Stokes I to Stokes Q are both included.}
\end{deluxetable*}

The ALMA interferometric data were calibrated with Common Astronomy Software Applications\footnote{https://casa.nrao.edu} (CASA, version: 5.4.0) following the calibration procedures described in \citet{2019PASP..131g5003G}. 
Since ALMA has linearly polarized feeds, a contribution from Stokes Q, U, and parallactic angle appears in the real part of all correlations (XX, YY, XY, YX). Here is a brief description of the polarization calibration procedures \citep{2019PASP..131g5003G}. We firstly obtained gain solutions for the polarization calibrator (3C279). Next, the X-Y cross-phase offset of the reference antenna and the Stokes Q, U parameters  for the polarization calibrator were determined simultaneously with the task {\tt gaincal} in {\tt XYf+QU} mode (3C279  has $\sim 130\arcdeg$ parallactic angle coverage).  Finally, we solved for the instrumental polarization  determining the leakage terms (or D-terms) for all antennas with the task {\tt polcal}.
The polarization leakage is of the order of a few percent and consistent across  frequency bands.

As for imaging, the CASA {\tt TCLEAN} task was applied to all Stokes parameters (I, Q, U and V) data for each subband to obtain polarization images of M87. 
In the {\tt TCLEAN} task, we set {\tt NTERMS=2}, {\tt CELL= 0.2$\arcsec$}, and a Briggs weighting with {\tt Robust=1}. We applied a primary beam correction and a manual mask that covers the regions from the core to the extended inner lobes of M87 in the {\tt TCLEAN} procedure. The subband images in Stokes I have rms noise levels of 2.5, 2.2, 2.0, and 2.0 $\rm mJy\ beam^{-1}$, and synthesized beams of $\rm 2\farcs9 \times 2\farcs3$, $\rm 2\farcs8 \times 2\farcs2$,$\rm 2\farcs5 \times 2\farcs1$, and $\rm 2\farcs4 \times 2\farcs0$, at 86.3, 88.3, 98.3, and 101.3 GHz, respectively. To obtain the image of linear polarized intensity, the CASA task {\tt IMMATH} was used with the {\tt POLI} mode. The linear polarized rms noise levels in each subband are 2.0, 2.0, 1.5, and 1.5 $\rm mJy\ beam^{-1}$ at 86.3, 88.3, 98.3, and 101.3 GHz, respectively.

We used the {\tt IMFIT} task to measure the polarized properties of the M87 core and the jet knots assuming elliptical Gaussian components. Table \ref{tab:knot} displays these measurements at 86.3 GHz (spw0). 
Defined by $ \chi= \rm \frac{1}{2}arctan(U/Q)$, the observed EVPA ($\chi_{\mathrm{obs}}$) was calculated with the {\tt IMMATH} task in the {\tt POLA} mode. The uncertainty of EVPA was calculated with the propagation of the standard error following the definition of $\chi$. Following \citet{2021ApJ...910L..14G}, the calculation also incorporates a 5\% systematic error caused by the absolute flux-scale calibration, and an additional 0.03\% systematic error due to leakage from Stokes I to Stokes Q and U \citep[corresponding to a minimum detectable polarization of 0.1\%,][]{Remijan2019}; systematic errors were included in the calculation in a quadratic form.

\begin{figure*}[ht!]
\plotone{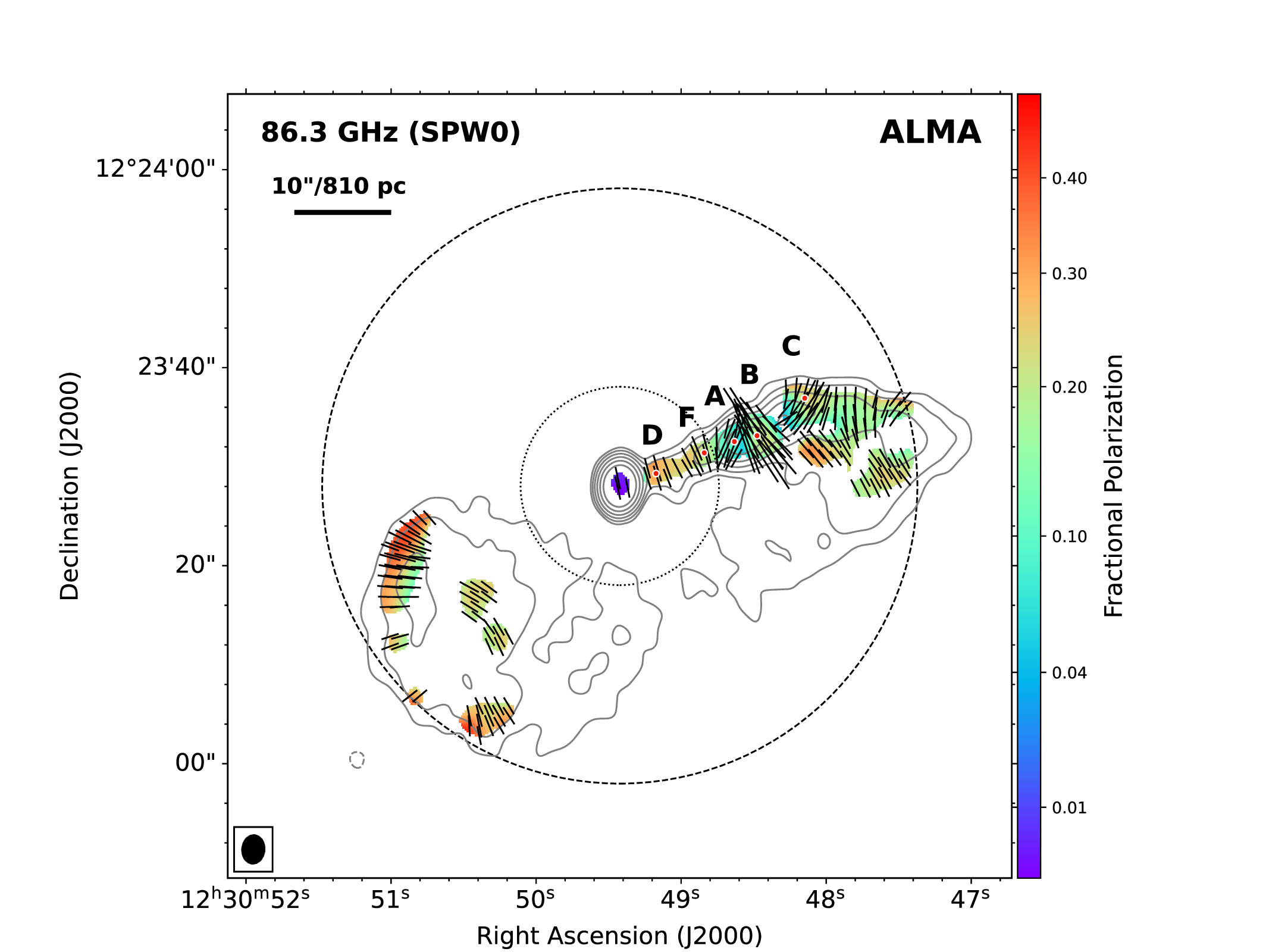}
\caption{\label{fig:m87} ALMA image of the continuum (contour) and linear polarization (color) image of the M87 central region at 86 GHz (SPW0), enclosing the core, the one-sided jet and the two inner lobes. The dashed line denotes the FWHM of the primary beam and the dotted line represents its inner 1/3 region of primary beam. The image center is at the M87 unresolved core. The gray contours starts at 4 times the stokes I rms noise ($\rm 1.4\ mJy\ beam^{-1}$) and increase in steps of two. The linear polarized electron vectors are marked by the black sticks and only shown for regions where the polarized intensity is greater than 3 times the polarized rms level.
The stick length is proportional to the polarized intensity. The synthesis beam, $\rm 2\farcs9 \times 2\farcs3,\ -4.5^{o}$, is  displayed as a filled ellipse at the bottom left corner. The prominent knots A, B, C and D are marked as red dots with white edges. We note that this image does not include systematic errors.}
\end{figure*}

To obtain the Faraday rotation measure (RM), clean images of the individual SPW were used. Firstly, we tapered the visibilities of SPWs 1-3 and restored image to match the lower resolution of SPW 0 that is $\rm 2\farcs9 \times 2\farcs3$ at $\rm P.A.= -4.5\arcdeg$ by setting the {\tt RESTORATION} parameters in the {\tt TCLEAN} task. Each SPW was then separately imaged with the above {\tt TCLEAN} parameters in full Stokes parameters. Finally, we applied the {\tt RMFIT} task to the individual SPW images to make a Faraday RM image and the associated error map with a polarized threshold of $\rm 6\ mJy\ beam^{-1}$ ($\rm 3\times rms$). No systematic errors were included in {\tt RMFIT}.
We note that, in principle, ALMA can only guarantee polarization information within the inner 1/3 of the primary beam\footnote{https://almascience.eso.org/documents-and-tools/cycle5/alma-technical-handbook}, i.e., approximately the inner $\rm 20\arcsec$ at Band 3. 
However, by checking the calibrators, we confirm that both on-axis ($\rm \leq 20\arcsec$) and off-axis ($\rm 20\sim 60\arcsec$) polarization leakages are below the threshold of our polarization analysis ($\rm 6\ mJy\ beam^{-1}$), which means the polarization leakage does not significantly influence the polarization results and Faraday RM images. 
The latter showcase ordered polarization structures along and across the entire jet (see Figs. \ref{fig:m87} and \ref{fig:rm}), providing further indication that  ALMA is capable of reliably recovering the polarized signal outside of the inner 1/3 beam  \citep[see also][]{2020PASP..132i4501H}.

\section{Results} \label{sec:results}

Figure \ref{fig:m87} shows the 86.3 GHz linearly polarized intensity image for the central $1\arcmin$ region of M87, where the compact core, the northwest jet, and the two inner lobes are evident. The total intensity (Stokes I) shown as gray contours illustrates the overall emission structure. 
The radio core dominates the Stokes I (total) flux with a peak brightness of $\rm 1.407\pm0.006 \ Jy\ beam^{-1}$
derived from CASA {\tt IMFIT} task.
The jet knots, including knots A, B, C, D and F, are more prominent in the linearly polarized intensity image than in the total intensity image. Specifically, the peak polarized flux density is found at knot B, with a peak flux density of $\rm 36.7\pm1.8 \ mJy\ beam^{-1}$. In contrast, previous optical and radio (centimeter) observations pinpointed knot A as the peak in polarized intensity \citep{1999AJ....117.2185P,2016ApJ...832....3A}.

The fractional polarization image, defined as the ratio of the linearly polarized intensity to the Stokes I intensity, is shown as the background color in Figure \ref{fig:m87}.
The lowest fractional polarization is found in the core at a level of only $\rm (0.88\pm 0.08)\%$. The fraction increases to $\rm (28.4\pm4.8)\%$ at knot D and gradually decreases downstream of the jet to only $\rm (6\pm0.6)\%$ at knot A and rebounds to $\rm (22.2\pm1.0)\%$ at knot B.
The two inner lobes show an overall higher fractional polarization than the jet. In particular, values up to $\rm (40\sim 50)\%$ are seen along the edge of the eastern inner lobe, spatially corresponding to the shock due to the termination of the counter jet \citep{1992Natur.355..802S, 1992Natur.355..804S}.

The black sticks in Figure \ref{fig:m87} represent the EVPA, the length of which corresponds to the polarization intensity.
The EVPA is highly aligned in the core and along the inner part of the jet out to knot F. The aligned EVPA between knot D to F indicates that the projected magnetic field, which is perpendicular to the EVPA, is well ordered and aligned with the jet.
The EVPA becomes significantly distorted further downstream in the jet, especially in knots A and B, where the polarized intensities are the highest among the jet knots. 
The EVPA distribution along the jet (and the corresponding magnetic field topology) is strikingly similar to that obtained with ALMA at 1.3~mm ($213-230$~GHz) by \citet[][e.g., their Figure 2]{2021ApJ...910L..14G}, including the substantial change of EVPA in knots A and B. 
The latter was also found in previous optical and radio polarimetric observations,
suggesting the existence of a shock in the immediate upstream of knot A \citep{2016ApJ...832....3A}.
The eastern inner lobe exhibits an overall ordered EVPA, i.e., EVPA is constant or changes smoothly and continuously over several beams. In particular, the inferred projected magnetic field appears to be well aligned with the edge of the eastern inner lobe.
The western lobe shows less ordered EVPA, and there is no significant enhancement of fractional polarization along its edge. This difference could be due to the different ambient medium and pressure between the eastern and western inner lobes. It is suggested that a molecular gas with a total mass of $\rm M_{H_2} \sim 4.7 \times 10^5\ M_{\odot}$ exists outside the eastern lobe \citep{2018MNRAS.475.3004S}. The interstellar pressure is also different as suggested by several X-ray cavities surrounding the eastern lobe and a huge shock cocoon on the north side of the western lobe \citep{2005xrrc.procE7.09K,2010MNRAS.407.2046M}. The linearly polarized intensity images of the other three sub-bands have the same structures as Figure \ref{fig:m87} (SPW0).

\begin{figure*}[ht]
\plotone{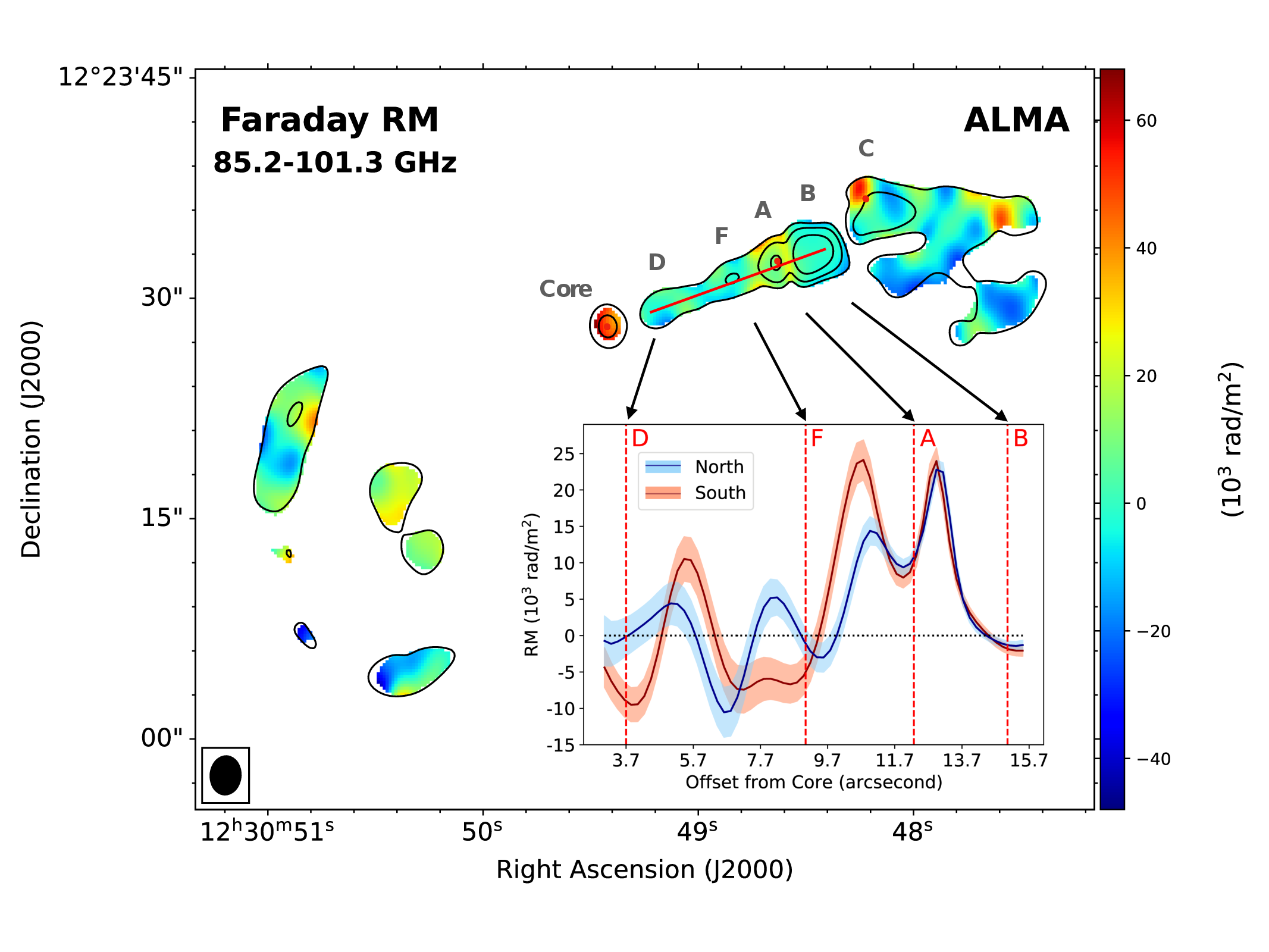}
\caption{\label{fig:rm} ALMA Faraday rotation measure of the core, jet and eastern inner lobe, overlaid with polarized intensity contours at levels of $ \rm(4, 8, 16, 32)\times\rm 1.4 \ mJy\ beam^{-1}$, the RMS level of the polarized intensity. The positions of knots D, F, A, B and C are marked as red dots with capital letters. The red dots mark the locations of the selected points, and their EVPA distributions at different frequencies are shown in Figure \ref{fig:evpa}.  The size of synthesis beam, $\rm 2\farcs9 \times 2\farcs3,\ -4.5^{o}$, is displayed as a filled ellipse at the bottom left corner. We note that this image does not include systematic errors. The insert shows RM variations on the northern (blue) and southern (red) side along the jet axis (red solid line), averaged over a width of $1\arcsec$. The dark red (blue) curve is encompassed by the orange (light blue) strip representing 1-$\sigma$ error. The red vertical dash lines mark the centroids of knots D, F, A, B and C. The horizontal black dotted line marks the zero level of the RM.
}
\end{figure*}

\begin{figure*} 
\plotone{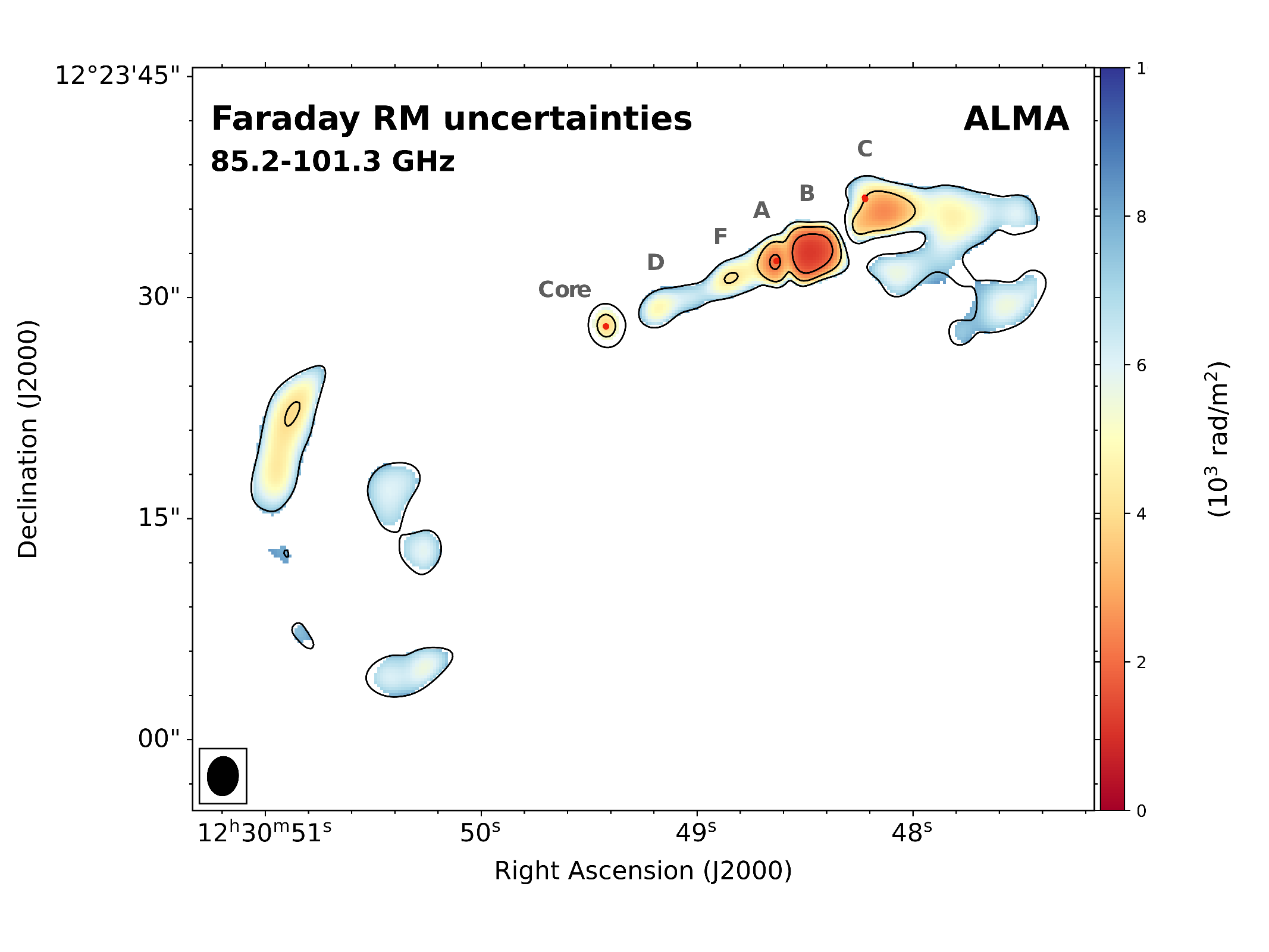}
\caption{\label{fig:rmerr} The uncertainties (one standard deviation) of Faraday rotation measure of the core, jet and eastern inner lobe, overlaid with polarized intensity contours at levels of $ \rm(4, 8, 16, 32)\times\rm 1.4 \ mJy\ beam^{-1}$, the RMS level of the polarized intensity. The positions of knots D, F, A, B and C are marked with capital letters.  The red dots mark the locations of the selected points, and their EVPA distributions at different frequencies are shown in Figure \ref{fig:evpa}. The size of synthesis beam, $\rm 2\farcs9 \times 2\farcs3,\ -4.5^{o}$, is displayed as a filled ellipse at the bottom left corner. We note that this image does not include systematic errors.} 
\end{figure*}

The spatial distribution of RM, as derived from the EVPA between 86.3 GHz to 100.3 GHz following Eqn.~\ref{eqn:RM}, is shown in Figure \ref{fig:rm}, and Figure \ref{fig:rmerr} displays the RM uncertainties\footnote{It is worth noting that both Figure \ref{fig:rm} and \ref{fig:rmerr} do not present any systematic errors, only the direct outputs from CASA task {\tt RMFIT}.}
The EVPA vector at the radio core is spatially well-ordered with a position angular range of $\rm (0\sim15)\arcdeg$ at a resolution of $\rm 2\arcsec$. The RM distribution of the core varies in the range of
$\rm (3.5 - 6.6)\times10^{4}\ rad\ m^{-2}$ and appears asymmetric,
showing on-average higher (lower) values at the southeastern (northwestern) side, which is probably due to Laing-Garrington effect where the jet closer to us appears brighter and more polarized than its counter-jet \citep{1988Natur.331..147G,1988Natur.331..149L}. The RM value at the core centroid ($[\rm R.A. , Decl.$] = $[12^{\mathrm{h}}30^{\mathrm{m}}49\fs423, +12\arcdeg23\arcmin28\farcs05]$) is $\rm (4.5\pm 0.4)\times10^{4}\ rad\ m^{-2}$. 
The average RM within a $1\arcsec$-radius circle, approximately equal to the synthesized beam size, is $\rm (4.9\pm0.9)\times10^{4}\ rad\ m^{-2}$. This asymmetric RM distribution with a low northwest side and a high southeast side is not an artificial pattern, because the distribution of RM errors in the core is radially symmetric. As for the the core shift contribution, it is only $\rm (3\pm12)$ micro-arcseconds for M87 core from 86.3 to 100.3 GHz \citep{2011Natur.477..185H}, which is significantly smaller than the synthesized beam size. 
The asymmetric RM distribution can be caused by the inner jet, including HST-1 \citep{1991AJ....101.1632B}, which is 0\farcs8 - 0\farcs9 offset from the central SMBH \citep{2007ApJ...663L..65C, 2012A&A...538L..10G}. 

\begin{figure*}[t]
\plotone{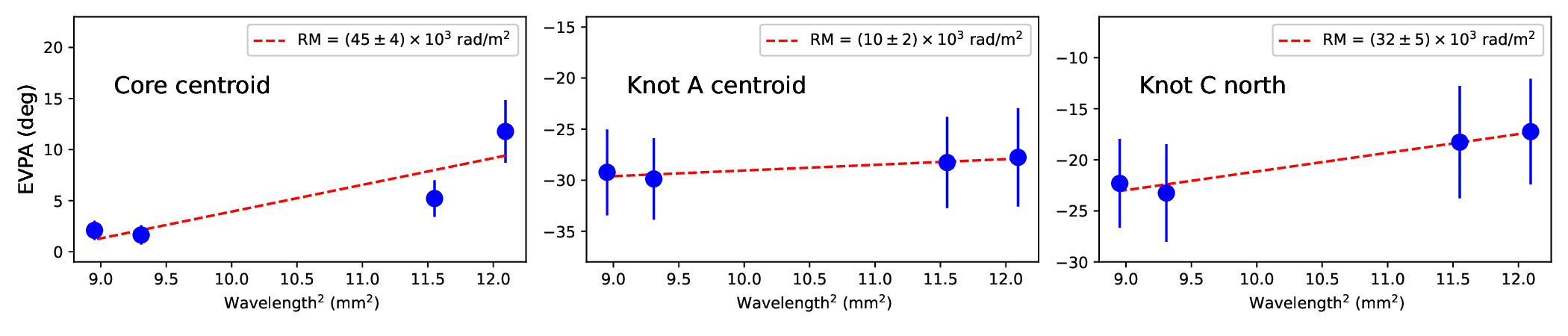}
\caption{The EVPAs versus $\rm \lambda^{2}$ of the core and the sampled position in the M87 jet knots. 
The best-fit slope is shown as a red dashed line and indicated in each panel. Data were taken at the red points in Figure \ref{fig:rm}. Panels from left to right show significant RM value, close to zero value, and moderate RM value, respectively. We note that the three panels contain thermal noise and the systematic errors mentioned in Section 2.\label{fig:evpa}} 
\end{figure*}

Along the jet from knot D to B, the RM value spans a wide range from $\sim -2\times10^4\rm~rad\ m^{-2}$ to $\sim 3\times10^4\rm~rad\ m^{-2}$ with a typical error of $0.3\times10^4\rm~rad\ m^{-2}$. 
Figure \ref{fig:evpa} plots the measured EVPAs of the core and the sampled position versus wavelength square of the four subbands. 
Due to the limited frequency coverage, we adopted a linear fitting to estimate the RM for both the core and the sampled position in the jet knots. The derived RM values are significantly different among the jet knots (Figure \ref{fig:rm}).
The insert panel in Figure \ref{fig:rm} illustrates the RM variations on the northern and southern side of the jet axis, as indicated by the red line. 
The RM curve of both sides is averaged over a width of $1\arcsec$ perpendicular to the jet axis. 
Between knots D and F,
the value of RM shows a sequential sign change on both the southern (red) and northern (blue) sides within a range of $\pm1.0\times10^4\rm~rad\ m^{-2}$. 
Downstream of knot F, the RM quickly increases on both sides to a peak value of $\sim2.5\times10^4\rm~rad\ m^{-2}$  about half-way to knot A and then decreases to a local minimum of $\sim0.8\times10^4\rm~rad\ m^{-2}$ at a distance of 11.7 arcseconds from the core, which is close to the centroid of knot A.
Further downstream, taking roughly the same pace on both sides, the RM first rises rapidly to another peak of $\sim2.3\times10^4\rm~rad\ m^{-2}$ immediately post-knot A.  A similar, albeit weaker, increasing in the RM in front of knot A was detected at $8 - 43$ GHz by \citet{2016ApJ...823...86A}, and the location is consistent with the so-called A-shock position \citep{2016ApJ...832....3A,1999AJ....117.2185P}. Then the RM drops steeply to a level of $(-0.03\pm 0.07)\times10^4\rm~rad\ m^{-2}$ at the centroid of knot B. 
The RM errors at knots A and B are relatively small due to their high linear polarized intensities. 

Towards the more extended structure to knot C, the RM rises again especially near the northeast side of knot C, where RM values up to $(5.5\pm0.6)\times10^{4}\rm\ rad\ m^{-2}$ are detected. 
Previous 6-cm observations confirmed an increase in RM at knot C, but at a rather lower value of only a few hundred $\rm rad\ m^{-2}$ \citep{1990ApJ...362..449O,2016ApJ...823...86A}. 
In the inner lobes, the RM varies from $\rm -2\times10^{4}$ to $\rm 5\times10^{4}\ rad\ m^{-2}$ with a typical error of $\rm 0.5\times10^{4}\ rad\ m^{-2}$. 
Previous VLA observations at 6 cm \citep{1990ApJ...362..449O} revealed that, on average, the absolute RM values in the inner lobe are higher than those in the jet.
We also found RM sign reversals in the inner lobes, which indicates a reversal of the magnetic field direction along the line of sight. This phenomenon has not been detected in previous centimeter observations \citep{1990ApJ...362..449O}.

\section{Discussion}

\subsection{Faraday rotation measure at the core}

The core exhibits an RM gradient roughly following the jet orientation, in the sense that the southeastern part is two times higher than the northwestern part (Figure \ref{fig:rm}). 
Due to the limited 1$\arcsec$ resolution of our observations, it is hard to determine whether such a gradient is due to HST-1 or to the inner mas-scale jet (see details in Section \ref{sec:results}).
However, since HST-1 is optically thin at millimeter wavelengths \citep{2010A&A...515A..38C}, 
the RM of the core could be dominated by the mas-scale jet or even the inner part. The gradual decrease electron density along the direction of the jet outward can cause the observed decrease in RM along the northwest direction.

There are two possible origins of the observed Faraday rotation: 1) an internal Faraday screen from an accretion flow 
\citep{2017MNRAS.468.2214M,2020MNRAS.498.5468R,2019ApJ...875L...1E}; 2) foreground external Faraday screens including the probable presence of a wind in the vicinity of a jet \citep{2019ApJ...871..257P,2022ApJ...924..124Y} or jet sheath \citep{2004ApJ...612..749Z,2016ApJ...823...86A,2020A&A...637L...6K}. 
Both the turbulence of the accretion flow and the varying external screen can cause the RM variation. In turn, the variability in RM can help constrain the physical size of the Faraday screen and/or the emitting source (assuming it is variable). For example, \citet{2021ApJ...910L..14G}
present a model where the RM variability at mm wavelengths in M87 can be explained by a rapidly varying source on horizon-scale and a static Faraday screen.

\begin{figure}[ht]
\plotone{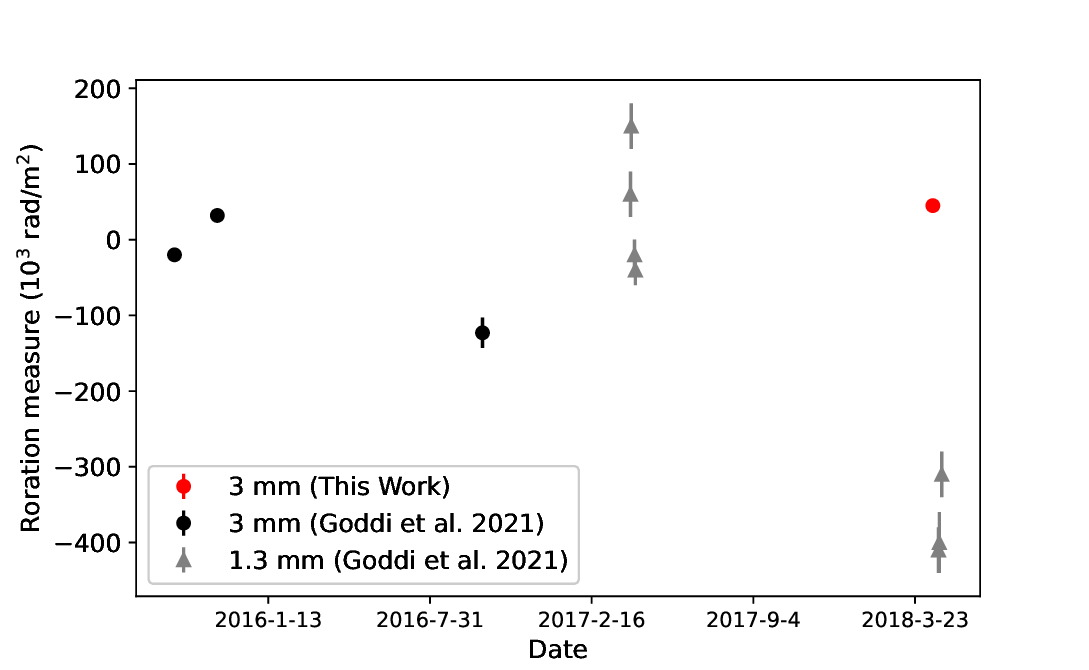}
\caption{The RM variation in the M87 core from 2015 to 2018\label{fig:sign} at 3 mm and 1.3 mm by ALMA observations. Our 3 mm measurement is denoted by the red circle, while the 3 mm and 1.3 mm measurements by \citet{2021ApJ...910L..14G} are denoted by the black circles and grey triangles, respectively. 
} 
\end{figure}

Figure \ref{fig:sign} shows RM measurements by ALMA, including this work and \citet{2019PASP..131g5003G, 2021ApJ...910L..14G} from 2016 to 2018 at 1.3 mm (Band 6) and 3.5 mm (Band 3). Despite the different wavelengths, the ALMA array configuration differences resulted in similar observational resolutions. The RM sign was found to reverse from negative (in 2016) to positive (in 2018), which suggests a change in the core RM sign at least twice over the 2.5-year period. 
A remarkable sign reversal occurred between April 14 and April 21, in 2018, within one week the positive RM of $(4.5\pm0.4)\times10^{4}\rm~rad\ m^{-2}$ at 3 mm changed to negative with a mean RM of $\rm (-4.1\pm0.3)\times10^{5}\ rad\ m^{-2}$ at 1.3 mm.
Such a short-term variability implies that the Faraday rotation took place in a very compact region (within 10 Schwarzschild radii, $R_{\rm sch}$) in the vicinity of the SMBH.
Both an internal Faraday screen introduced by turbulence of accretion flows \citep{2017MNRAS.468.2214M,2021ApJ...910L..13E} and/or a rapid and compact emitting source \citep{2021ApJ...910L..14G} can be used to explain the RM variability with sign reversal on the weekly scale.
We caution that the 1.3 mm and 3.5 mm emissions may not arise from the same location and hence might be subject to a substantially different configuration of the local magnetic field. 

In addition to the RM variability, the high ($\rm \sim 10^4\ rad\ m^{-2}$) RM detected in our millimeter observations supports a scenario where the Faraday rotation at the core occurs near the SMBH. 
Previous polarization sensitive VLBI observations at centimeter wavelengths have probed the RM properties inside the M87 Bondi radius with milliarcsecond resolution 
For example, \citet{2019ApJ...871..257P} found that the absolute RM value at $2-5$ GHz increases toward the SMBH, and the value exceeds 1000 $\rm rad\ m^{-2}$ at a distance of less than $\rm 10^{4}$ $R_{Sch}$ from the SMBH. They proposed a wind generated by hot accretion flows as an external Faraday screen, and predicted that the RM would be $\rm 2\times 10^{4}\ rad\ m^{-2}$ at a de-projected distance from the SMBH of about $\rm 2\times 10^{3}$ $R_{Sch}$. According to this model, our RM measurement of $\sim \rm 10^{4}\ rad\ m^{-2}$ could come from a location several thousand $R_{Sch}$ away from the SMBH.

\subsection{Faraday rotation measure of the jet on arc-second scales}

As shown in Figure \ref{fig:rm}, RM sign reversals occur both in the direction parallel to the jet and in the direction perpendicular to the jet from knot D to B. Since the M87 jet has a small viewing angle \citep[$\rm \sim 17\arcdeg$,][]{2009MNRAS.395..301W}, a change in the pitch angle of the helical magnetic field in the jet can cause RM sign reversals in a direction parallel to the jet.
Such RM sign reversals across the jet from the knot D to F can be interpreted as a result of the toroidal component of a helical magnetic field in the jet, which has been observed in a number of other AGN jets, e.g., 3C273 \citep{2002PASJ...54L..39A} and 1226+023 \citep{2012AJ....144..105H}. Using VLA observations at $4-18$ GHz, \citet{2021ApJ...923L...5P} found an RM reversal between the knots E and F. We note that this region is also where the transverse RM gradient is most significant in our 3 mm results. We also note that the 3 mm RM signs of the northern and southern parts are opposite to their 4-18 GHz result, which may be due to limited resolution and time variability.

As for the knots A and B, polarization fraction, EVPA and RM change drastically in the upstream of both knots (10.7 and 12.0 arcseconds from the M87 core). This implies that the magnetic field structure is significantly bent upstream of both knots A and B, where in-situ magnetic shocks occur (e.g., quad relativistic magnetohydrodynamics shocks \citep{2010ApJ...721.1783N}). 
As a possible consequence of magnetic field bending, the helical structure of the magnetic field is broken at the knot A and the coil is unwound at knot B \citep{2010ApJ...721.1783N,2013ApJ...774L..21M}. This scenario can explain why the 3.5 mm RM of knots A and B have a similar pattern on the northern and southern sides.  
Furthermore, RM sign reversals of the knot C and the western lobe suggest that the helical structure of magnetic field may slightly remain in these places, supporting the above scenario.

To better understand the jet structure, it would be beneficial to compare our RM measurements at $86-101$\,GHz with those at other wavelengths.
At the lower frequencies of $8-43$\,GHz, VLA observations revealed  RM values of the jet knots at a level of several hundred $\rm rad\ m^{-2}$ from the knot D to knot C \citep{2016ApJ...823...86A}. This is two orders of magnitude lower than our RM result, although the resolution of our observations ($\sim 2\arcsec$) is lower than their VLA observations ($\sim 0.5\arcsec$). Assuming the observed RM at different frequencies comes from the same radiation mechanism and that the surrounding environment has a stable magnetic field and a continuous material, such a difference in RM values at different frequencies suggests that the M87 jet is stratified \citep{1999AJ....117.2185P} from millimeter to centimeter wavelengths.

The RM-frequency relationship can be used to estimate the jet geometry. Because of the optical depth effects ($\rm \tau \sim 1$), the thickness of the screen through which the lower-frequency photons pass is smaller.
\citet{2007AJ....134..799J} measured the RM of 15 highly variable AGN jets. Assuming a helical magnetic field, an at least mildly relativistic Faraday screen, and a power-law decreasing gradient in the electron density of the screen, they derived a RM-frequency dependence of $\rm |RM| \propto \nu^{a}$. $a$ is related to the profile of a jet electron density ($P_{n_e}$). The jet density profile can be described as a function of distance $r$ from the black hole, $P_{n_e} \propto r^{-a}$. For example, $a =2 $ means that the Faraday rotation is occurring in a conically expanding jet, while lower values imply a more highly collimated jet \citep[e.g.,][]{2009MNRAS.393..429O}.
This relationship can be applied to estimate the geometry of M87 jet. We adopt the range $\rm |RM| = (0-250)\ rad\ m^{-2}$ (with a median of 150 $\rm rad\ m^{-2}$) in the frequency range $4 -18$ GHz \citep{2021ApJ...923L...5P}. For the frequency range $8-43$ GHz, we use the range $\rm |RM| = (0-500)\ rad\ m^{-2}$ (with a median of 250 $ \rm  rad\ m^{-2}$) between knot A and D 
from \citet{2016ApJ...823...86A}.
For the frequency range $86-101$ GHz, we use our ALMA result range $\rm RM = (1-2.5)\times 10^4\ rad\ m^{-2}$. Using an orthogonal distance regression from 4 to 101 GHz, we obtain $\rm \alpha=2.3\pm0.4$ for the kpc-scale jet from the knot A to knot D. Our calculated alpha value indicates a trumpet jet shape with an increasingly larger opening angle as the distance from the SMBH increases. This shape corresponds to an overly expanding jet, which likely to happen in a decelerating jet \citep{2014MNRAS.437.3405L}.

\section{Summary}
We have studied the 3 mm ALMA polarization images of the M87 jet. The total intensity and linear polarized intensity are consistent with previous studies \citep{2021ApJ...910L..14G}. 
The average RM value of the core was $\rm (4.5\pm 0.4)\times10^{4}\ rad\ m^{-2}$ on April 14/15, 2018. 
The M87 jet showcases RM values of the order of tens of thousands $\rm rad\ m^{-2}$ with sign reversals. Time-variability and an extreme value of the RM suggest that the Faraday screen at the core is internal, although a rapidly variable  emitting (compact) source could also explain the observed RM.
The RM gradient in the jet is consistent with a well-ordered helical magnetic field at the kpc scale \citep{2010ApJ...721.1783N,2016ApJ...832....3A}. 
A comparison between the 3 mm RM value and measurements at longer wavelengths \citep{2016ApJ...823...86A,2021ApJ...923L...5P} suggests that the kpc-scale jet between knot A and D has a frequency dependence in RM and that the jet may propagate in a trumpet-like shape.

\acknowledgments
{We thank Prof. Dr. Eduardo Ros, Dr. Jongho Park, and the anonymous referee for helpful suggestions. R.L. is supported by the National Science Fund for Distinguished Young Scholars of China (Grant No. 12325302), the Key Program of the National Natural Science Foundation of China (grant no. 11933007), the Key Research Program of Frontier Sciences, CAS (grant no. ZDBS-LY-SLH011), the Shanghai Pilot Program for Basic Research, Chinese Academy of Sciences, Shanghai Branch (JCYJ-SHFY-2021-013) and the Max Planck Partner Group of the MPG and the CAS. C.G. acknowledges support by FAPESP (Fundação de Amparo à Pesquisa do Estado de São Paulo) under grant 2021/01183-8. I.M.-V. acknowledges fundings support from Projects PID2022-140888NB-C22 and PID2019-108995GB-C22 (Ministerio de Ciencia, Innovacion y Universidades), and Project ASFAE/2022/018 (Generalitat Valenciana). J.Y.K. is supported for this research by the National Research Foundation of Korea (NRF) grant funded by the Korean government (Ministry of Science and ICT; grant no. 2022R1C1C1005255).  Z.L. acknowledges support by the National Natural Science Foundation of China (grant 12225302) R.-Y. L. acknowledges the National Natural Science Foundation of China under the grant No. 12393852. 

This paper makes use of the following ALMA data: ADS/JAO.ALMA\#2017.1.00842.V. ALMA is a partnership of ESO (representing its member states), NSF (USA) and NINS (Japan), together with NRC (Canada), MOST and ASIAA (Taiwan), and KASI (Republic of Korea), in cooperation with the Republic of Chile. The Joint ALMA Observatory is operated by ESO, AUI/NRAO and NAOJ. }

\vspace{5mm}
\facility{ALMA}
\software{APLpy \citep{2012ascl.soft08017R},  
    CASA \citep{2022PASP..134k4501C}.}

\clearpage
\bibliography{M87}{}
\bibliographystyle{aasjournal}

\end{document}